\documentclass[aps,prl,reprint,groupedaddress]{revtex4-1}

\usepackage{amsmath}
\usepackage{graphicx}
\usepackage{dcolumn}
\usepackage{bm}

\newcommand{\fig}[4][t]{\begin{figure}[#1]\begin{center}\includegraphics[scale=#2]{#3.pdf}\vspace{-0.25 cm}\caption{#4}\label{fig:#3}\end{center}\end{figure}}

\begin{document}

\title{Continuous parametric feedback cooling of a single atom in an optical cavity}

\author{C. Sames}
\author{C. Hamsen}
\author{H. Chibani}
\altaffiliation[present address: ]{Physics department, King Fahd University of Petroleum and Minerals, Dhahran 34463, Saudi Arabia}
\author{P. A. Altin}
\email[email: ]{paul.altin@anu.edu.au}
\altaffiliation[present address: ]{Australian National University, Canberra, Australian Capital Territory 0200, Australia}
\author{T. Wilk}
\author{G. Rempe}
\affiliation{Max-Planck-Institut f\"ur Quantenoptik, Hans-Kopfermann-Str.\ 1, D-85748 Garching, Germany}

\date{\today}

\begin{abstract}
We demonstrate a new feedback algorithm to cool a single neutral atom trapped inside a standing-wave optical cavity. The algorithm is based on parametric modulation of the confining potential at twice the natural oscillation frequency of the atom, in combination with fast and repetitive atomic position measurements. The latter serve to continuously adjust the modulation phase to a value for which parametric excitation of the atomic motion is avoided. Cooling is limited by the measurement back action which decoheres the atomic motion after only a few oscillations. Nonetheless, applying this feedback scheme to a $\sim5$ kHz oscillation mode increases the average storage time of a single atom in the cavity by a factor of 60 to more than 2 seconds. In contrast to previous feedback schemes, our algorithm is also capable of cooling a much faster $\sim500$ kHz oscillation mode within just microseconds. This demonstrates that parametric cooling is a powerful technique that can be applied in all experiments where optical access is limited.
\end{abstract}


\maketitle

\section{Introduction}
As in classical devices, measurement-based feedback control is expected to become an indispensable tool for stabilization and protection of individual quantum systems against decoherence~\cite{Dowling:2003, Giovannetti:2011, Escher:2011}. But quantum feedback control is fundamentally different to its classical counterpart, due to the inevitable backaction caused by a measurement~\cite{Braginsky:1980, Braginsky:1992, Wiseman:1994, Lloyd:2000, Nelson:2000, Smith:2002, Steck:2004, Gillett:2010, Sayrin:2011, Vijay:2012}. The latter is famously exemplified in Heisenberg's microscope thought experiment~\cite{Heisenberg:1949} that demonstrates how position measurements perturb the momentum of a particle and thus heat the motional degrees of freedom. Here we investigate parametric feedback for cooling a single neutral atom in an optical cavity in the regime where the quantum backaction from the measurement easily disturbs the atomic motion. Our technique makes use of position information extracted from a probe laser transmitted through the cavity, and requires only a single control laser along the cavity axis to actuate on the atomic motion, making it ideally suited to miniaturized systems with poor optical access for additional cooling beams. Cooling atoms in cavities is not only of interest for investigating quantum feedback and control but also improves the localization of the atom in the cavity mode and thus the coupling strength, which is crucial for the observation of a variety of fundamental quantum effects~\cite{Birnbaum:2005, Schuster:2008, Kubanek:2008, Ourjoumtsev:2011}. In contrast to previously demonstrated feedback schemes that rely on single-shot or discrete algorithms~\cite{Fischer:2002, Kubanek:2009, Koch:2010}, the technique described here employs a continuous parametric feedback strategy. Effectively, this increases the duty cycle, decreases the required measurement rate, and can thus cool the fast ($\sim$ 500\,kHz) oscillation of the atom along the axis of a standing-wave cavity field. We also demonstrate an increase of the average storage time of a single atom in the cavity by a factor of 60 to more than 2 seconds.

Parametric excitation of large numbers of particles in a confining potential is familiar in cold atom physics, where it is known as a major source of heating in far off-resonance optical traps~\cite{Savard:1997, Gehm:1998} and is often used to measure trapping frequencies~\cite{Friebel:1998}. In addition, it has been demonstrated that parametric excitation in an anharmonic trap can selectively evaporate atoms with higher energies, leading to cooling~\cite{Poli:2002}. Parametric resonance phenomena have also been studied in micro- and nano-electromechanical systems~\cite{Turner:1998}, and parametric feedback has been used to stabilize nano-mechanical oscillators~\cite{Villanueva:2011} and to generate squeezing~\cite{Vinante:2013}.

Continuous parametric cooling of a single particle by active feedback was recently investigated~\cite{Gieseler:2012} and cooling of a laser-trapped nanoparticle to a center-of-mass temperature of less than 1\,mK was demonstrated~\cite{Jain:2016}. However, in these experiments the particle undergoing feedback was much more massive than a single atom, and therefore remained virtually undisturbed by the measurement. Weak probing and long integration times reduce the measurement-induced back action enabling sub-Doppler cooling of a single ion in a Paul trap through direct feedback~\cite{Bushev:2006}. However, in contrast to an ion that is held in a trapping potential that exceeds its kinetic energy by orders of magnitude, the atom in our experiment is held in a much shallower optical potential that exceeds its kinetic energy only by about a factor of two~\cite{Koch:2010}. Therefore, the atom is very sensitive to the measurement-induced back action and its motion decoheres after only a few oscillations in the trap.

Our feedback technique relies crucially on a continuous, weak probing of the atomic position, from which we update the parametric feedback amplitude and phase on a timescale comparable to the oscillation period. The system is in the strong-coupling regime of cavity quantum electrodynamics (QED), where the cavity resonance splits into two normal modes and the transmission at the empty-cavity resonance is reduced to near zero. As the atom moves away from the center of the cavity mode, the coupling strength is reduced and the transmission increases. Continuous modulation of the trap potential with an amplitude and phase determined by the measured probe transmission removes kinetic energy from the atom. Note that the velocity-dependent cooling force originates from the periodic modulation of the conservative force confining the atom, not from two subsequent position measurements that are made to determine the velocity of the atom~\cite{Fischer:2002, Kubanek:2009, Koch:2010}. As a matter of principle, this allows one to reduce the measurement rate and therefore the associated measurement-induced back action from the probe beam. In fact, cooling in our experiment is achieved with a probe beam intensity so low that the resulting photon detection rate is smaller than or at most on the same order as the trap frequency, either $\sim$~5\,kHz or $\sim$~500\,kHz.

\section{Experimental setup}

\fig{1.00}{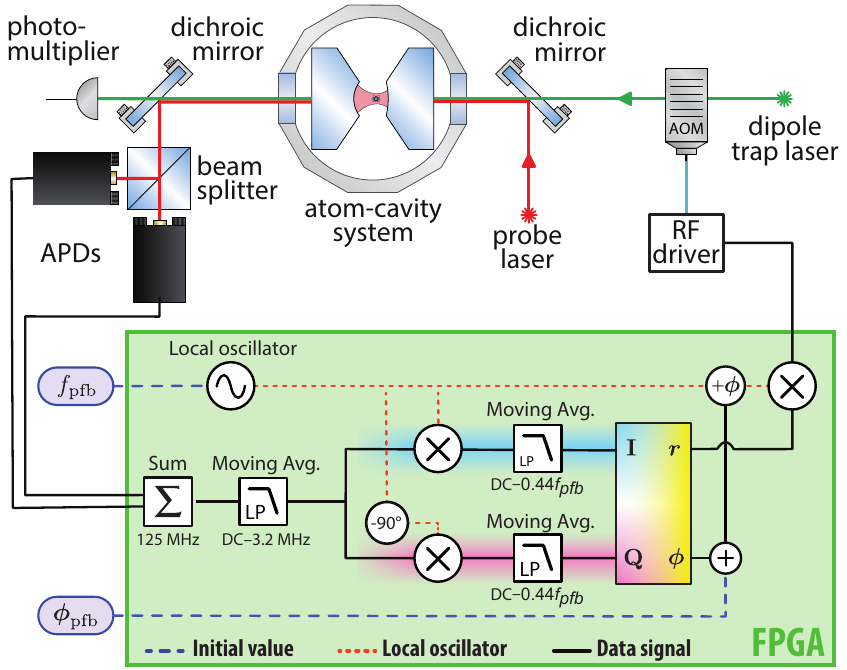}{(Color online.) Experimental setup. A single atom is trapped inside the optical cavity by a red-detuned dipole trap. This system is probed with a resonant beam that is detected by avalanche photo diodes (APDs) in photon counting mode and demodulated in real-time at a preset frequency $f_{\text{pfb}}$ on a field-programmable gate array (FPGA) shown in the green box. The amplitude and phase are extracted and used to feedback to the atom via a sinusoidal intensity modulation of the dipole trap. The parameter $\phi_{\text{pfb}}$ advances the phase of the feedback with respect to the atom's oscillation to enable cooling.}

Our cavity QED apparatus has been described in detail elsewhere \cite{Koch:2010,Koch:2011}. Briefly, single $^{85}$Rb atoms launched from an atomic fountain are captured in a standing wave dipole trap inside a high-finesse ($\mathcal{F} = 195,000$) Fabry-Perot cavity of length 260\,$\mu$m (Figure \ref{fig:figure1}). The maximal atom-cavity coupling constant $g/2\pi = 16$\,MHz exceeds both the atomic dipole decay rate ($\gamma/2\pi = 3.0$\,MHz) and the field decay rate of the cavity ($\kappa/2\pi = 1.5$\,MHz), placing our system in the strong coupling regime of cavity QED $g \gg (\gamma,\kappa)$. The dipole trap beam is red-detuned by four free-spectral ranges from the rubidium D$_2$ line at $\lambda=780$\,nm, and also serves to stabilize the cavity length. The intracavity dipole trap power is typically 950\,nW, creating a trap with depth $k_B \times 850\,\mu$K and harmonic oscillation frequencies $\omega_\rho/2\pi = 4.8$\,kHz and $\omega_z/2\pi = 500\,$kHz in the radial and axial directions, respectively.

Our weak measurement of the atom's position in the cavity comes from the transmission of a weak ($\sim 3$\,pW) TEM$_{00}$ probe beam which is close to resonance with the $5S_{1/2}, F=3\rightarrow5P_{3/2},F=4$ cycling transition. As a consequence of the normal-mode splitting, the probe transmission at empty cavity resonance drops monotonically as the coupling strength increases \cite{Boca:2004,Maunz:2005,Kubanek:2011}. This strength, however, varies as a function of position due to the cavity mode function, which has a Gaussian profile in the radial direction and varies sinusoidally with a period of $\lambda/2$ along the cavity axis due to the standing wave pattern of the probe beam. The cavity transmission therefore directly depends on the position of the atom in the cavity mode.

\fig{1.0}{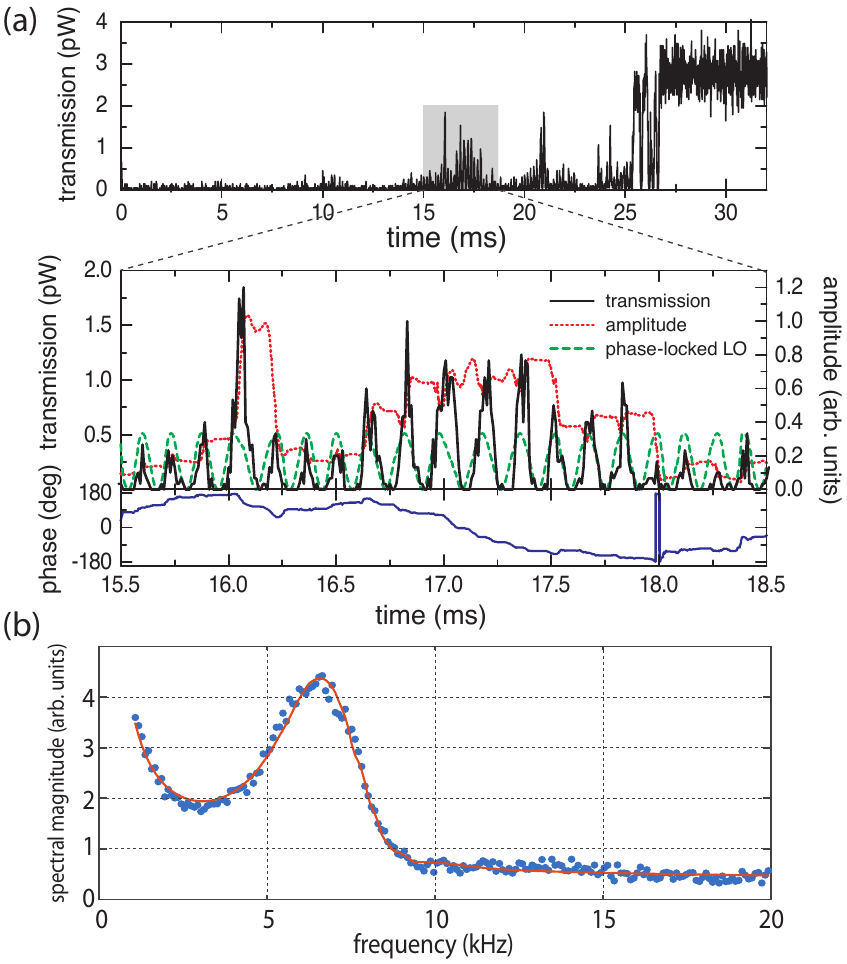}{(Color online.) Observing the atomic motion. (a) Transmission for a typical atom storage event; low transmission indicates the presence of an atom. The zoomed-in section shows how the FPGA locks to the atomic oscillation. The black (solid) trace is the measured cavity transmission, the red (dotted) is related to the inferred atomic oscillation amplitude (see text), and the green (dashed) is the local oscillator phase-locked to the measured atomic oscillation. Below in blue (solid) is shown the measured phase difference between the local oscillator and the atomic motion. (b) Fourier transform of measured transmission. The peak appears below the expected frequency of $2\omega_\rho/2\pi$ due to the nonlinear dependence of cavity transmission on radial position. It is broadened due to the measurement back action, reflecting decoherence of the atomic oscillation by the probing laser beam. A fit incorporating the nonlinear transmission function, which connects the atomic position to the transmitted intensity, yields a $Q$-factor of 2.8 (see text for details).}

As the atom oscillates back and forth in the trapping potential, the coupling constant and thus the cavity transmission vary at twice the oscillation frequency. The transmitted probe signal is demodulated at approximately this frequency, and the amplitude and phase of the motion are fed back via modulation of the intracavity dipole trap intensity in order to cool the atom. An example feedback trace is shown in Figure \ref{fig:figure2}(a). The detected amplitude is used to adjust the modulation strength, which allows us to damp the feedback for very cold atoms which yield insufficient phase information to suppress heating. Continual updating of the modulation phase based on the information obtained from the measurement is critical to the success of this method. If the phase were fixed, the oscillating atom would phase-lock to the parametric drive and its energy would increase exponentially, eventually expelling it from the trap. To avoid this, phase of the feedback signal is adjusted according to the measured phase difference between the local oscillator and the atomic motion (blue trace in Fig. \ref{fig:figure2}a) plus a variable phase advance $\phi_\text{pfb}$.

The price paid for this measurement, of course, is that the atom can also scatter probe photons into free space modes outside of the cavity. The random nature of this spontaneous emission leads to decoherence of the atom's oscillatory motion and heating. In our experiment, the scattering rate during radial feedback was roughly 20\,kHz, which is sufficient to decohere atomic oscillations in the dipole trap within a few periods. Our feedback loop therefore operates in the regime of poor quality factor, where a measurement can only be used to predict the future motion of the atom for a short time.

We measure the quality factor of the oscillation by fitting the Fourier transform of the cavity transmission (Fig.\ \ref{fig:figure2}b). This analysis needs to take into account the nonlinear atom-cavity coupling function and the anharmonicity of the trapping potential. The cavity transmission relates nonlinearly to the actual (radial) atomic position via the cavity coupling function $g(r) \sim e^{-r^2/w_0^2}$. Additionally, the anharmonic potential causes the atomic oscillation frequency to depend on amplitude (higher amplitudes result in lower frequencies), which skews the transmission peak towards lower frequencies. We fit the measured transmission using a damped harmonic oscillator model that incorporates these effects, as well as a $1/f$ noise background from the photodetectors. This is shown as the solid line in Fig.\ \ref{fig:figure2}b, and yields a quality factor of $Q = 2.8\pm0.1$.

The cavity transmission is measured by four avalanche photodiodes (APDs) in photon counting mode and sent to a field-programmable gate array, which also runs the feedback algorithm (Fig.\ \ref{fig:figure1}). The FPGA output controls an AOM which modulates the intensity of the dipole trap laser (and thus the trap stiffness) at frequency $f_{\textrm{pfb}}$, with a phase and amplitude determined by the measured atomic motion.

The feedback algorithm implemented on the FPGA is shown schematically in Fig.\ \ref{fig:figure1}. It begins by summing the number of photons detected during 8\,ns time intervals. This signal is sent through a moving average low-pass filter with a 3\,dB bandwidth of 3.2\,MHz, and the in-phase and out-of-phase quadrature components are extracted by multiplying it with a local oscillator. These two channels represent the complex oscillation amplitudes at the feedback frequency $f_{\text{pfb}}$. These amplitudes are integrated by another moving average filter, with the integration time $\tau$ adjusted to an integer multiple of the oscillation period, $\tau = n / f_{\text{pfb}}$. The moving average therefore acts as a low-pass filter with a 3\,dB bandwidth of $n \times 0.44f_{\text{pfb}}$. Non-integer multiples will lead to artefacts, especially when $\tau$ is on the order of only a few oscillation periods. As this filtering is applied in the complex plane, it acts as a resolution bandwidth filter (RBW) allowing to extract the narrow-band spectral component at $f_{\text{pfb}}$.

A subsequent complex-to-polar conversion extracts the magnitude and phase information from the complex oscillation amplitude. Note that the inferred magnitude (red curve in Fig. \ref{fig:figure2}a) represents the amplitude of the atomic oscillation as expressed by the cavity transmission, which relates nonlinearly to the actual atomic position via the cavity coupling function (see above). The phase is used to lock the local oscillator generated by the FPGA to the detected oscillation. The phase of the oscillator is additionally advanced by the parameter $\phi_{\text{pfb}}$, which allows us to tune the feedback algorithm to either heat or cool the atom. The amplitude of the output modulation is varied depending on the inferred magnitude of the atomic oscillation.

\section{Radial feedback results}

The performance of the feedback algorithm is determined by measuring the average time for which a captured atom remains in the trap before escaping. Since capture attempts are not always successful, we require that an atom remains in the cavity mode for at least 2\,ms (corresponding to $\sim10$ radial oscillation periods) for it to be considered trapped. Storage times are averaged over at least 100 individual measurements of a trapped atom.

The optimum modulation amplitude was found empirically to be 11\% for the maximum inferred atomic oscillation (that which causes the transmission to increase to its empty-cavity value), corresponding to an increase in the trap depth by $k_B \times 90\,\mu$K. A typical modulation during feedback was 6\%. The optimum integration time was found to be one oscillation period, $\tau = 1/f_{\text{pfb}}$. This is expected due to the low $Q$-factor: the measurement back action renders the atomic motion incoherent over timescales longer than the oscillation period, so integrating for longer than this degrades our estimate of the atom's amplitude and phase.

\fig{1.0}{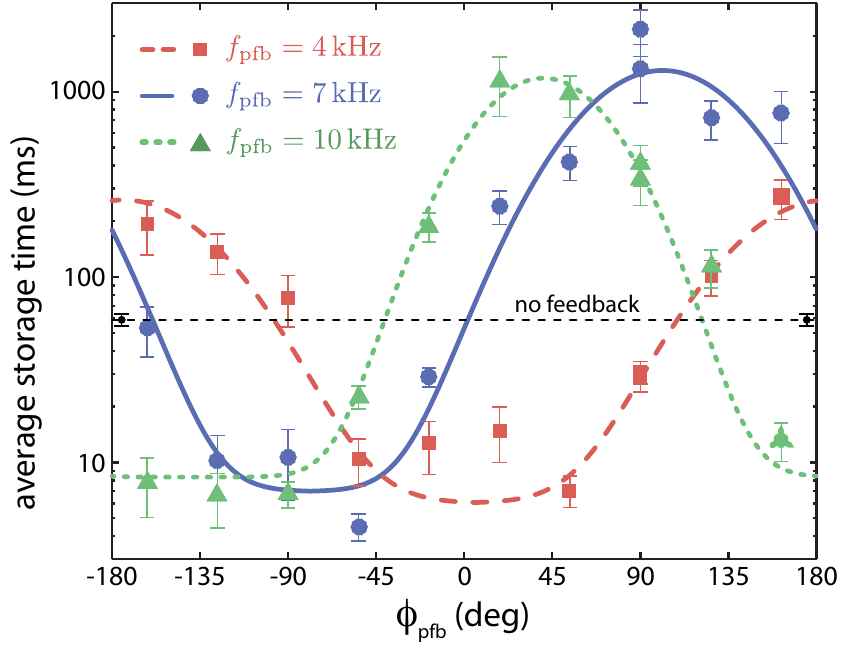}{(Color online.) Radial parametric feedback. Average storage time of an atom in the cavity as a function of phase advance $\phi_{\text{pfb}}$ for various feedback frequencies. The phase advance which gives optimal cooling decreases as the feedback frequency increases. Error bars represent the standard error in the mean for the storage time. The lines are fits of a periodic Gaussian that is used to determine the ideal phase advance. The dashed horizontal line shows the average storage time without feedback.}

Figure \ref{fig:figure3} shows the average storage time with feedback for single atoms trapped in the cavity, as a function of the phase advance $\phi_{\text{pfb}}$ and for $f_{\text{pfb}} = 4$, 7 and 10\,kHz. These frequencies are too slow to interact with the axial atomic motion and therefore affect only the radial oscillation, whose natural frequency is 4.8\,kHz. Nonetheless, the feedback has a strong effect on the storage times, which for $f_{\text{pfb}} = 7$\,kHz vary by more than two orders of magnitude as $\phi_{\text{pfb}}$ is tuned. Both cooling and heating are observed at different values of the phase advance, and the phase advance which results in the best cooling performance decreases as the feedback frequency is increased. The maximum average storage time is approximately 2\,s at $f_{\text{pfb}} = 7$\,kHz, and occurs at a phase advance of approximately $\phi_{\text{pfb}} = \pi/2$. By contrast, the average storage time without feedback is $(55 \pm 5)$\,ms.

The qualitative behavior observed is as expected for a parametrically driven oscillator. When the parametric modulation is out of phase with the oscillator's motion, the potential relaxes when the oscillator is at the center of its swing and stiffens as it reaches the turning points, thus amplifying its motion. When the modulation is in phase, on the other hand, the situation is reversed and the oscillator's motion is damped. Normally, near a parametric resonance, the oscillator phase-locks to the drive and absorbs energy exponentially. However, our active feedback algorithm maintains the phase difference between the measured atomic motion and the parametric drive, preventing phase locking and allowing us to parametrically cool as well as heat.

\fig[b]{1.0}{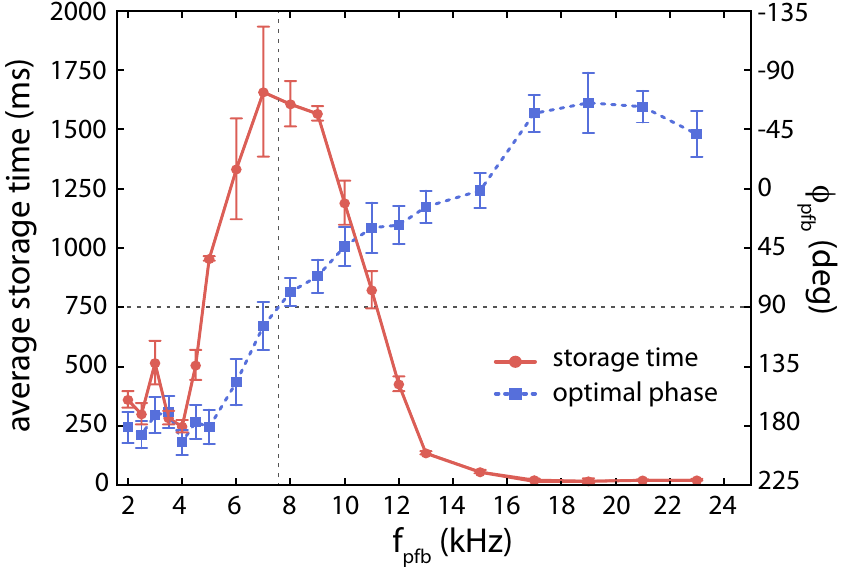}{(Color online.) Radial parametric feedback versus feedback frequency. Measured average storage time (circles, solid line) and optimal phase advance (squares, dotted line) as a function of feedback frequency $f_{\text{pfb}}$. The error bars represent the standard error in the mean for the storage time, and the fitting error in the determination of optimal phase.}

We also investigate the performance of the feedback algorithm as a function of the parametric driving frequency $f_{\text{pfb}}$. The maximum average storage time, obtained from fits such as those shown in Fig.\ \ref{fig:figure3}, is plotted as a function of $f_{\text{pfb}}$ in Fig.\ 4, as well as the phase advance at which the corresponding maximum was measured. The feedback performance peaks at around $f_{\text{pfb}} = 7$\,kHz before falling off at higher frequencies. The optimal phase advance crosses $\pi/2$ at the peak in the cooling performance. The observed behavior is due to a combination of the detection procedure and the parametric oscillator response. The oscillator response is maximum at the principal parametric resonance at $2\omega_\rho/2\pi$, with smaller peaks at the higher order resonances $2\omega_\rho /2\pi n$. The in-phase/quadrature (IQ) demodulation, with a moving average filter whose kernel length is exactly one oscillation period, results in a broad sinc-shaped envelope which shifts the peak lower in frequency. Longer integration times would give a sharper peak closer to the parametric resonance.

\section{Axial feedback results}

Previous attempts to apply feedback cooling to the fast oscillation of an atom along a standing wave dipole trap have failed~\cite{Kubanek:2009,Koch:2010}, since not enough information about the atom's position could be obtained within a single oscillation period to effect useful feedback. Here, we show that our new technique can also be applied to cool the atom's motion along the cavity axis, which due to the standing wave trapping potential has a much higher natural oscillation frequency of approximately 500\,kHz. This is possible because the feedback algorithm exploits our pre-knowledge of the oscillation frequency (both in demodulating the transmission signal at $2\omega_z/2\pi$ and in modulating the trap stiffness at this frequency), and thus relies less on information extracted from the real-time measurement of cavity transmission.

\fig{1.0}{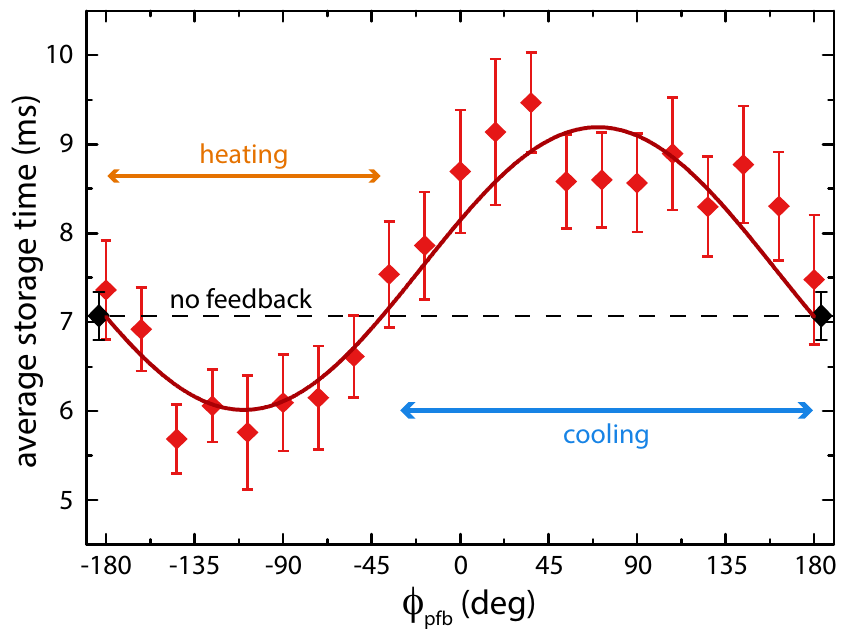}{(Color online.) Axial parametric feedback. Performance of axial feedback cooling at $f_{\text{pfb}} = 625$\,kHz, corresponding to twice the natural oscillation frequency of the atom along the axis of the standing wave dipole trap. The data points represent experimental measurements of storage time averaged over at least 100 runs per point, and the solid line is a sinusoidal fit to the data. Error bars represent the standard error in the mean for the storage time.}

Figure \ref{fig:figure5} shows the average storage time as a function of phase advance $\phi_{\text{pfb}}$ for a feedback loop operated at $f_{\text{pfb}} = 625$\,kHz, which was found empirically to give the best feedback performance. The modulation amplitude was 36\% for the maximum inferred atomic oscillation. (Note that the phase advance $\phi_{\text{pfb}}$ here is not absolute, as small delays in the signal path are not negligible at these frequencies.) As in Fig.\ \ref{fig:figure3}, it is seen that the algorithm is capable of either cooling or heating the atom, depending on whether the parametric feedback is kept in phase or out of phase with the axial oscillation. At the optimal phase advance, the average storage time shows an improvement of more than 30\% compared to the case without feedback. Note that the storage time without feedback, $(7.1\pm0.3)$\,ms, is smaller than in Fig.\ \ref{fig:figure3} due to the fact that the probe power was increased by a factor of 10 compared to the radial feedback experiments, which is necessary to ensure the algorithm has enough information to operate at the higher frequency. The success of axial cooling demonstrates the speed and versatility of our parametric feedback algorithm. We note that the atoms are expected to move less in the axial direction since cavity cooling acts along this axis, which may explain the relative effectiveness of radial feedback over axial feedback.

\section{Conclusions and Outlook}

We have demonstrated a continuous parametric feedback algorithm that efficiently cools a single atom held in a standing wave dipole trap within an optical cavity. The back action caused by our measurement of the atom's position decoheres its oscillatory motion within a few periods; nonetheless, by applying feedback to the radial motion of the atom, its average storage time can be increased by a factor of 60 to 2\,s. Additionally, the feedback algorithm is able to damp the motion of the atom in either the radial direction or along the cavity axis, where it oscillates at a frequency of about 500\,kHz.

In addition to experiments on measurement-based quantum feedback, our work has immediate impact on fundamental investigations of quantum-optical phenomena and application-oriented research in quantum-information processing. This comes from the fact that cooling improves the localization of the atom in the cavity mode and thus the atom-cavity coupling strength $g$. With all parameters of the atom and the cavity alone kept constant, the increase of $g$ has (at least) two distinct advantages: First, it directly increases the photon-number dependent splitting between the dressed states of the strongly coupled atom-cavity system. This makes it easier to selectively drive single- or even multi-photon transitions between certain pairs of these states. As a result, genuine quantum-nonlinear phenomena like photon blockade \cite{Birnbaum:2005,Hamsen:2017} and conditional phase shifts \cite{Turchette:1995,Beck:2016} become more pronounced. Second, a stronger atom-cavity coupling enhances the photon emission and absorption rate of an atom into and from the cavity, respectively, as compared to the corresponding rates in free space \cite{Northup:2014,Reiserer:2015}. This is a reminiscence of the Purcell effect and makes the atom more one-dimensional from a radiation point of view. It dramatically improves the efficiencies of unidirectional atom-photon quantum devices including entangled-qubit sources \cite{Wilk:2007,Weber:2009,Stute:2012} and single-qubit memories \cite{Specht:2011,Korber:2018}. In addition to these advantages, cooling not only increases the average value of $g$ but also decreases the thermal fluctuations of $g$. This makes it easier to discriminate between classical and quantum noise and possibly even eliminate the former completely. Last but not least, the fact that our technique uses only information extracted from the cavity transmission makes it ideally suited to miniaturized systems with large atom-cavity coupling but without optical access for additional cooling beams \cite{Dayan:2008,Gehr:2010,Rosenblum:2015,Scheucher:2016}.

\section{Acknowledgements}

\begin{acknowledgments}
C.S.\ acknowledges financial support from the Bavarian Ph.D.\ program of excellence QCCC, P.A.\ from the Alexander von Humboldt Foundation and the European Union through the Marie Curie Initial Training Network CCQED (Project No. 264666), and C.H. from the Deutsche Forschungsgemeinschaft via the excellence cluster Nanosystems Initiative Munich (NIM).
\end{acknowledgments}

\bibliography{parametric}

\begin{thebibliography}{49}%
\makeatletter
\providecommand \@ifxundefined [1]{%
 \@ifx{#1\undefined}
}%
\providecommand \@ifnum [1]{%
 \ifnum #1\expandafter \@firstoftwo
 \else \expandafter \@secondoftwo
 \fi
}%
\providecommand \@ifx [1]{%
 \ifx #1\expandafter \@firstoftwo
 \else \expandafter \@secondoftwo
 \fi
}%
\providecommand \natexlab [1]{#1}%
\providecommand \enquote  [1]{``#1''}%
\providecommand \bibnamefont  [1]{#1}%
\providecommand \bibfnamefont [1]{#1}%
\providecommand \citenamefont [1]{#1}%
\providecommand \href@noop [0]{\@secondoftwo}%
\providecommand \href [0]{\begingroup \@sanitize@url \@href}%
\providecommand \@href[1]{\@@startlink{#1}\@@href}%
\providecommand \@@href[1]{\endgroup#1\@@endlink}%
\providecommand \@sanitize@url [0]{\catcode `\\12\catcode `\$12\catcode
  `\&12\catcode `\#12\catcode `\^12\catcode `\_12\catcode `\%12\relax}%
\providecommand \@@startlink[1]{}%
\providecommand \@@endlink[0]{}%
\providecommand \url  [0]{\begingroup\@sanitize@url \@url }%
\providecommand \@url [1]{\endgroup\@href {#1}{\urlprefix }}%
\providecommand \urlprefix  [0]{URL }%
\providecommand \Eprint [0]{\href }%
\providecommand \doibase [0]{http://dx.doi.org/}%
\providecommand \selectlanguage [0]{\@gobble}%
\providecommand \bibinfo  [0]{\@secondoftwo}%
\providecommand \bibfield  [0]{\@secondoftwo}%
\providecommand \translation [1]{[#1]}%
\providecommand \BibitemOpen [0]{}%
\providecommand \bibitemStop [0]{}%
\providecommand \bibitemNoStop [0]{.\EOS\space}%
\providecommand \EOS [0]{\spacefactor3000\relax}%
\providecommand \BibitemShut  [1]{\csname bibitem#1\endcsname}%
\let\auto@bib@innerbib\@empty
\bibitem [{\citenamefont {Jonathan P.~Dowling}(2003)}]{Dowling:2003}%
  \BibitemOpen
  \bibfield  {author} {\bibinfo {author} {\bibfnamefont {G.~J.~M.}\
  \bibnamefont {Jonathan P.~Dowling}},\ }\href
  {http://www.jstor.org/stable/3559215} {\bibfield  {journal} {\bibinfo
  {journal} {Philosophical Transactions: Mathematical, Physical and Engineering
  Sciences}\ }\textbf {\bibinfo {volume} {361}},\ \bibinfo {pages} {1655}
  (\bibinfo {year} {2003})}\BibitemShut {NoStop}%
\bibitem [{\citenamefont {Giovannetti}\ \emph {et~al.}(2011)\citenamefont
  {Giovannetti}, \citenamefont {Lloyd},\ and\ \citenamefont
  {Maccone}}]{Giovannetti:2011}%
  \BibitemOpen
  \bibfield  {author} {\bibinfo {author} {\bibfnamefont {V.}~\bibnamefont
  {Giovannetti}}, \bibinfo {author} {\bibfnamefont {S.}~\bibnamefont {Lloyd}},
  \ and\ \bibinfo {author} {\bibfnamefont {L.}~\bibnamefont {Maccone}},\ }\href
  {http://dx.doi.org/10.1038/nphoton.2011.35} {\bibfield  {journal} {\bibinfo
  {journal} {Nat. Photon.}\ }\textbf {\bibinfo {volume} {5}},\ \bibinfo {pages}
  {222} (\bibinfo {year} {2011})}\BibitemShut {NoStop}%
\bibitem [{\citenamefont {Escher}\ \emph {et~al.}(2011)\citenamefont {Escher},
  \citenamefont {de~Matos~Filho},\ and\ \citenamefont
  {Davidovich}}]{Escher:2011}%
  \BibitemOpen
  \bibfield  {author} {\bibinfo {author} {\bibfnamefont {B.~M.}\ \bibnamefont
  {Escher}}, \bibinfo {author} {\bibfnamefont {R.~L.}\ \bibnamefont
  {de~Matos~Filho}}, \ and\ \bibinfo {author} {\bibfnamefont {L.}~\bibnamefont
  {Davidovich}},\ }\href {http://dx.doi.org/10.1038/nphys1958} {\bibfield
  {journal} {\bibinfo  {journal} {Nat. Phys.}\ }\textbf {\bibinfo {volume}
  {7}},\ \bibinfo {pages} {406} (\bibinfo {year} {2011})}\BibitemShut {NoStop}%
\bibitem [{\citenamefont {Vladimir B.~Braginsky}(1980)}]{Braginsky:1980}%
  \BibitemOpen
  \bibfield  {author} {\bibinfo {author} {\bibfnamefont {K.~S.~T.}\
  \bibnamefont {Vladimir B.~Braginsky}, \bibfnamefont {Yuri I.~Vorontsov}},\
  }\href {\doibase 10.1126/science.209.4456.547} {\bibfield  {journal}
  {\bibinfo  {journal} {Science}\ }\textbf {\bibinfo {volume} {209}},\ \bibinfo
  {pages} {547} (\bibinfo {year} {1980})}\BibitemShut {NoStop}%
\bibitem [{\citenamefont {Braginsky}\ and\ \citenamefont
  {Khalili}(1992)}]{Braginsky:1992}%
  \BibitemOpen
  \bibfield  {author} {\bibinfo {author} {\bibfnamefont {V.~B.}\ \bibnamefont
  {Braginsky}}\ and\ \bibinfo {author} {\bibfnamefont {F.~Y.}\ \bibnamefont
  {Khalili}},\ }\href@noop {} {\emph {\bibinfo {title} {Quantum Measurement}}}\
  (\bibinfo  {publisher} {Cambridge Univ. Press},\ \bibinfo {address}
  {Cambridge},\ \bibinfo {year} {1992})\BibitemShut {NoStop}%
\bibitem [{\citenamefont {Wiseman}(1994)}]{Wiseman:1994}%
  \BibitemOpen
  \bibfield  {author} {\bibinfo {author} {\bibfnamefont {H.~M.}\ \bibnamefont
  {Wiseman}},\ }\href {\doibase 10.1103/PhysRevA.49.2133} {\bibfield  {journal}
  {\bibinfo  {journal} {Phys. Rev. A}\ }\textbf {\bibinfo {volume} {49}},\
  \bibinfo {pages} {2133} (\bibinfo {year} {1994})}\BibitemShut {NoStop}%
\bibitem [{\citenamefont {Lloyd}(2000)}]{Lloyd:2000}%
  \BibitemOpen
  \bibfield  {author} {\bibinfo {author} {\bibfnamefont {S.}~\bibnamefont
  {Lloyd}},\ }\href {\doibase 10.1103/PhysRevA.62.022108} {\bibfield  {journal}
  {\bibinfo  {journal} {Phys. Rev. A}\ }\textbf {\bibinfo {volume} {62}},\
  \bibinfo {pages} {022108} (\bibinfo {year} {2000})}\BibitemShut {NoStop}%
\bibitem [{\citenamefont {Nelson}\ \emph {et~al.}(2000)\citenamefont {Nelson},
  \citenamefont {Weinstein}, \citenamefont {Cory},\ and\ \citenamefont
  {Lloyd}}]{Nelson:2000}%
  \BibitemOpen
  \bibfield  {author} {\bibinfo {author} {\bibfnamefont {R.~J.}\ \bibnamefont
  {Nelson}}, \bibinfo {author} {\bibfnamefont {Y.}~\bibnamefont {Weinstein}},
  \bibinfo {author} {\bibfnamefont {D.}~\bibnamefont {Cory}}, \ and\ \bibinfo
  {author} {\bibfnamefont {S.}~\bibnamefont {Lloyd}},\ }\href {\doibase
  10.1103/PhysRevLett.85.3045} {\bibfield  {journal} {\bibinfo  {journal}
  {Phys. Rev. Lett.}\ }\textbf {\bibinfo {volume} {85}},\ \bibinfo {pages}
  {3045} (\bibinfo {year} {2000})}\BibitemShut {NoStop}%
\bibitem [{\citenamefont {Smith}\ \emph {et~al.}(2002)\citenamefont {Smith},
  \citenamefont {Reiner}, \citenamefont {Orozco}, \citenamefont {Kuhr},\ and\
  \citenamefont {Wiseman}}]{Smith:2002}%
  \BibitemOpen
  \bibfield  {author} {\bibinfo {author} {\bibfnamefont {W.~P.}\ \bibnamefont
  {Smith}}, \bibinfo {author} {\bibfnamefont {J.~E.}\ \bibnamefont {Reiner}},
  \bibinfo {author} {\bibfnamefont {L.~A.}\ \bibnamefont {Orozco}}, \bibinfo
  {author} {\bibfnamefont {S.}~\bibnamefont {Kuhr}}, \ and\ \bibinfo {author}
  {\bibfnamefont {H.~M.}\ \bibnamefont {Wiseman}},\ }\href {\doibase
  10.1103/PhysRevLett.89.133601} {\bibfield  {journal} {\bibinfo  {journal}
  {Phys. Rev. Lett.}\ }\textbf {\bibinfo {volume} {89}},\ \bibinfo {pages}
  {133601} (\bibinfo {year} {2002})}\BibitemShut {NoStop}%
\bibitem [{\citenamefont {Steck}\ \emph {et~al.}(2004)\citenamefont {Steck},
  \citenamefont {Jacobs}, \citenamefont {Mabuchi}, \citenamefont
  {Bhattacharya},\ and\ \citenamefont {Habib}}]{Steck:2004}%
  \BibitemOpen
  \bibfield  {author} {\bibinfo {author} {\bibfnamefont {D.~A.}\ \bibnamefont
  {Steck}}, \bibinfo {author} {\bibfnamefont {K.}~\bibnamefont {Jacobs}},
  \bibinfo {author} {\bibfnamefont {H.}~\bibnamefont {Mabuchi}}, \bibinfo
  {author} {\bibfnamefont {T.}~\bibnamefont {Bhattacharya}}, \ and\ \bibinfo
  {author} {\bibfnamefont {S.}~\bibnamefont {Habib}},\ }\href {\doibase
  10.1103/PhysRevLett.92.223004} {\bibfield  {journal} {\bibinfo  {journal}
  {Phys. Rev. Lett.}\ }\textbf {\bibinfo {volume} {92}},\ \bibinfo {pages}
  {223004} (\bibinfo {year} {2004})}\BibitemShut {NoStop}%
\bibitem [{\citenamefont {Gillett}\ \emph {et~al.}(2010)\citenamefont
  {Gillett}, \citenamefont {Dalton}, \citenamefont {Lanyon}, \citenamefont
  {Almeida}, \citenamefont {Barbieri}, \citenamefont {Pryde}, \citenamefont
  {O'Brien}, \citenamefont {Resch}, \citenamefont {Bartlett},\ and\
  \citenamefont {White}}]{Gillett:2010}%
  \BibitemOpen
  \bibfield  {author} {\bibinfo {author} {\bibfnamefont {G.~G.}\ \bibnamefont
  {Gillett}}, \bibinfo {author} {\bibfnamefont {R.~B.}\ \bibnamefont {Dalton}},
  \bibinfo {author} {\bibfnamefont {B.~P.}\ \bibnamefont {Lanyon}}, \bibinfo
  {author} {\bibfnamefont {M.~P.}\ \bibnamefont {Almeida}}, \bibinfo {author}
  {\bibfnamefont {M.}~\bibnamefont {Barbieri}}, \bibinfo {author}
  {\bibfnamefont {G.~J.}\ \bibnamefont {Pryde}}, \bibinfo {author}
  {\bibfnamefont {J.~L.}\ \bibnamefont {O'Brien}}, \bibinfo {author}
  {\bibfnamefont {K.~J.}\ \bibnamefont {Resch}}, \bibinfo {author}
  {\bibfnamefont {S.~D.}\ \bibnamefont {Bartlett}}, \ and\ \bibinfo {author}
  {\bibfnamefont {A.~G.}\ \bibnamefont {White}},\ }\href {\doibase
  10.1103/PhysRevLett.104.080503} {\bibfield  {journal} {\bibinfo  {journal}
  {Phys. Rev. Lett.}\ }\textbf {\bibinfo {volume} {104}},\ \bibinfo {pages}
  {080503} (\bibinfo {year} {2010})}\BibitemShut {NoStop}%
\bibitem [{\citenamefont {Sayrin}\ \emph {et~al.}(2011)\citenamefont {Sayrin},
  \citenamefont {Dotsenko}, \citenamefont {Zhou}, \citenamefont {Peaudecerf},
  \citenamefont {Rybarczyk}, \citenamefont {Gleyzes}, \citenamefont {Rouchon},
  \citenamefont {Mirrahimi}, \citenamefont {Amini}, \citenamefont {Brune},
  \citenamefont {Raimond},\ and\ \citenamefont {Haroche}}]{Sayrin:2011}%
  \BibitemOpen
  \bibfield  {author} {\bibinfo {author} {\bibfnamefont {C.}~\bibnamefont
  {Sayrin}}, \bibinfo {author} {\bibfnamefont {I.}~\bibnamefont {Dotsenko}},
  \bibinfo {author} {\bibfnamefont {X.}~\bibnamefont {Zhou}}, \bibinfo {author}
  {\bibfnamefont {B.}~\bibnamefont {Peaudecerf}}, \bibinfo {author}
  {\bibfnamefont {T.}~\bibnamefont {Rybarczyk}}, \bibinfo {author}
  {\bibfnamefont {S.}~\bibnamefont {Gleyzes}}, \bibinfo {author} {\bibfnamefont
  {P.}~\bibnamefont {Rouchon}}, \bibinfo {author} {\bibfnamefont
  {M.}~\bibnamefont {Mirrahimi}}, \bibinfo {author} {\bibfnamefont
  {H.}~\bibnamefont {Amini}}, \bibinfo {author} {\bibfnamefont
  {M.}~\bibnamefont {Brune}}, \bibinfo {author} {\bibfnamefont {J.-M.}\
  \bibnamefont {Raimond}}, \ and\ \bibinfo {author} {\bibfnamefont
  {S.}~\bibnamefont {Haroche}},\ }\href {http://dx.doi.org/10.1038/nature10376}
  {\bibfield  {journal} {\bibinfo  {journal} {Nature}\ }\textbf {\bibinfo
  {volume} {477}},\ \bibinfo {pages} {73} (\bibinfo {year} {2011})}\BibitemShut
  {NoStop}%
\bibitem [{\citenamefont {Vijay}\ \emph {et~al.}(2012)\citenamefont {Vijay},
  \citenamefont {Macklin}, \citenamefont {Slichter}, \citenamefont {Weber},
  \citenamefont {Murch}, \citenamefont {Naik}, \citenamefont {Korotkov},\ and\
  \citenamefont {Siddiqi}}]{Vijay:2012}%
  \BibitemOpen
  \bibfield  {author} {\bibinfo {author} {\bibfnamefont {R.}~\bibnamefont
  {Vijay}}, \bibinfo {author} {\bibfnamefont {C.}~\bibnamefont {Macklin}},
  \bibinfo {author} {\bibfnamefont {D.~H.}\ \bibnamefont {Slichter}}, \bibinfo
  {author} {\bibfnamefont {S.~J.}\ \bibnamefont {Weber}}, \bibinfo {author}
  {\bibfnamefont {K.~W.}\ \bibnamefont {Murch}}, \bibinfo {author}
  {\bibfnamefont {R.}~\bibnamefont {Naik}}, \bibinfo {author} {\bibfnamefont
  {A.~N.}\ \bibnamefont {Korotkov}}, \ and\ \bibinfo {author} {\bibfnamefont
  {I.}~\bibnamefont {Siddiqi}},\ }\href {http://dx.doi.org/10.1038/nature11505}
  {\bibfield  {journal} {\bibinfo  {journal} {Nature}\ }\textbf {\bibinfo
  {volume} {490}},\ \bibinfo {pages} {77} (\bibinfo {year} {2012})}\BibitemShut
  {NoStop}%
\bibitem [{\citenamefont {Heisenberg}(1949)}]{Heisenberg:1949}%
  \BibitemOpen
  \bibfield  {author} {\bibinfo {author} {\bibfnamefont {W.}~\bibnamefont
  {Heisenberg}},\ }\href@noop {} {\emph {\bibinfo {title} {The Physical
  Principles of the Quantum Theory}}}\ (\bibinfo  {publisher} {Courier Dover
  Publications},\ \bibinfo {year} {1949})\BibitemShut {NoStop}%
\bibitem [{\citenamefont {Birnbaum}\ \emph {et~al.}(2005)\citenamefont
  {Birnbaum}, \citenamefont {Boca}, \citenamefont {Miller}, \citenamefont
  {Boozer}, \citenamefont {Northup},\ and\ \citenamefont
  {Kimble}}]{Birnbaum:2005}%
  \BibitemOpen
  \bibfield  {author} {\bibinfo {author} {\bibfnamefont {K.~M.}\ \bibnamefont
  {Birnbaum}}, \bibinfo {author} {\bibfnamefont {A.}~\bibnamefont {Boca}},
  \bibinfo {author} {\bibfnamefont {R.}~\bibnamefont {Miller}}, \bibinfo
  {author} {\bibfnamefont {A.~D.}\ \bibnamefont {Boozer}}, \bibinfo {author}
  {\bibfnamefont {T.~E.}\ \bibnamefont {Northup}}, \ and\ \bibinfo {author}
  {\bibfnamefont {H.~J.}\ \bibnamefont {Kimble}},\ }\href
  {http://dx.doi.org/10.1038/nature03804} {\bibfield  {journal} {\bibinfo
  {journal} {Nature}\ }\textbf {\bibinfo {volume} {436}},\ \bibinfo {pages}
  {87} (\bibinfo {year} {2005})}\BibitemShut {NoStop}%
\bibitem [{\citenamefont {Schuster}\ \emph {et~al.}(2008)\citenamefont
  {Schuster}, \citenamefont {Kubanek}, \citenamefont {Fuhrmanek}, \citenamefont
  {Puppe}, \citenamefont {Pinkse}, \citenamefont {Murr},\ and\ \citenamefont
  {Rempe}}]{Schuster:2008}%
  \BibitemOpen
  \bibfield  {author} {\bibinfo {author} {\bibfnamefont {I.}~\bibnamefont
  {Schuster}}, \bibinfo {author} {\bibfnamefont {A.}~\bibnamefont {Kubanek}},
  \bibinfo {author} {\bibfnamefont {A.}~\bibnamefont {Fuhrmanek}}, \bibinfo
  {author} {\bibfnamefont {T.}~\bibnamefont {Puppe}}, \bibinfo {author}
  {\bibfnamefont {P.~W.~H.}\ \bibnamefont {Pinkse}}, \bibinfo {author}
  {\bibfnamefont {K.}~\bibnamefont {Murr}}, \ and\ \bibinfo {author}
  {\bibfnamefont {G.}~\bibnamefont {Rempe}},\ }\href
  {http://dx.doi.org/10.1038/nphys940} {\bibfield  {journal} {\bibinfo
  {journal} {Nat. Phys.}\ }\textbf {\bibinfo {volume} {4}},\ \bibinfo {pages}
  {382} (\bibinfo {year} {2008})}\BibitemShut {NoStop}%
\bibitem [{\citenamefont {Kubanek}\ \emph {et~al.}(2008)\citenamefont
  {Kubanek}, \citenamefont {Ourjoumtsev}, \citenamefont {Schuster},
  \citenamefont {Koch}, \citenamefont {Pinkse}, \citenamefont {Murr},\ and\
  \citenamefont {Rempe}}]{Kubanek:2008}%
  \BibitemOpen
  \bibfield  {author} {\bibinfo {author} {\bibfnamefont {A.}~\bibnamefont
  {Kubanek}}, \bibinfo {author} {\bibfnamefont {A.}~\bibnamefont
  {Ourjoumtsev}}, \bibinfo {author} {\bibfnamefont {I.}~\bibnamefont
  {Schuster}}, \bibinfo {author} {\bibfnamefont {M.}~\bibnamefont {Koch}},
  \bibinfo {author} {\bibfnamefont {P.~W.~H.}\ \bibnamefont {Pinkse}}, \bibinfo
  {author} {\bibfnamefont {K.}~\bibnamefont {Murr}}, \ and\ \bibinfo {author}
  {\bibfnamefont {G.}~\bibnamefont {Rempe}},\ }\href {\doibase
  10.1103/PhysRevLett.101.203602} {\bibfield  {journal} {\bibinfo  {journal}
  {Phys. Rev. Lett.}\ }\textbf {\bibinfo {volume} {101}},\ \bibinfo {pages}
  {203602} (\bibinfo {year} {2008})}\BibitemShut {NoStop}%
\bibitem [{\citenamefont {Ourjoumtsev}\ \emph {et~al.}(2011)\citenamefont
  {Ourjoumtsev}, \citenamefont {Kubanek}, \citenamefont {Koch}, \citenamefont
  {Sames}, \citenamefont {Pinkse}, \citenamefont {Rempe},\ and\ \citenamefont
  {Murr}}]{Ourjoumtsev:2011}%
  \BibitemOpen
  \bibfield  {author} {\bibinfo {author} {\bibfnamefont {A.}~\bibnamefont
  {Ourjoumtsev}}, \bibinfo {author} {\bibfnamefont {A.}~\bibnamefont
  {Kubanek}}, \bibinfo {author} {\bibfnamefont {M.}~\bibnamefont {Koch}},
  \bibinfo {author} {\bibfnamefont {C.}~\bibnamefont {Sames}}, \bibinfo
  {author} {\bibfnamefont {P.~W.~H.}\ \bibnamefont {Pinkse}}, \bibinfo {author}
  {\bibfnamefont {G.}~\bibnamefont {Rempe}}, \ and\ \bibinfo {author}
  {\bibfnamefont {K.}~\bibnamefont {Murr}},\ }\href {\doibase
  10.1038/nature10170} {\bibfield  {journal} {\bibinfo  {journal} {Nature}\
  }\textbf {\bibinfo {volume} {474}},\ \bibinfo {pages} {623} (\bibinfo {year}
  {2011})}\BibitemShut {NoStop}%
\bibitem [{\citenamefont {Fischer}\ \emph {et~al.}(2002)\citenamefont
  {Fischer}, \citenamefont {Maunz}, \citenamefont {Pinkse}, \citenamefont
  {Puppe},\ and\ \citenamefont {Rempe}}]{Fischer:2002}%
  \BibitemOpen
  \bibfield  {author} {\bibinfo {author} {\bibfnamefont {T.}~\bibnamefont
  {Fischer}}, \bibinfo {author} {\bibfnamefont {P.}~\bibnamefont {Maunz}},
  \bibinfo {author} {\bibfnamefont {P.}~\bibnamefont {Pinkse}}, \bibinfo
  {author} {\bibfnamefont {T.}~\bibnamefont {Puppe}}, \ and\ \bibinfo {author}
  {\bibfnamefont {G.}~\bibnamefont {Rempe}},\ }\href {\doibase
  10.1103/PhysRevLett.88.163002} {\bibfield  {journal} {\bibinfo  {journal}
  {Phys. Rev. Lett.}\ }\textbf {\bibinfo {volume} {88}},\ \bibinfo {pages}
  {163002} (\bibinfo {year} {2002})}\BibitemShut {NoStop}%
\bibitem [{\citenamefont {Kubanek}\ \emph {et~al.}(2009)\citenamefont
  {Kubanek}, \citenamefont {Koch}, \citenamefont {Sames}, \citenamefont
  {Ourjoumtsev}, \citenamefont {Pinkse}, \citenamefont {Murr},\ and\
  \citenamefont {Rempe}}]{Kubanek:2009}%
  \BibitemOpen
  \bibfield  {author} {\bibinfo {author} {\bibfnamefont {A.}~\bibnamefont
  {Kubanek}}, \bibinfo {author} {\bibfnamefont {M.}~\bibnamefont {Koch}},
  \bibinfo {author} {\bibfnamefont {C.}~\bibnamefont {Sames}}, \bibinfo
  {author} {\bibfnamefont {A.}~\bibnamefont {Ourjoumtsev}}, \bibinfo {author}
  {\bibfnamefont {P.~W.~H.}\ \bibnamefont {Pinkse}}, \bibinfo {author}
  {\bibfnamefont {K.}~\bibnamefont {Murr}}, \ and\ \bibinfo {author}
  {\bibfnamefont {G.}~\bibnamefont {Rempe}},\ }\href {\doibase
  10.1038/nature08563} {\bibfield  {journal} {\bibinfo  {journal} {Nature}\
  }\textbf {\bibinfo {volume} {462}},\ \bibinfo {pages} {898} (\bibinfo {year}
  {2009})}\BibitemShut {NoStop}%
\bibitem [{\citenamefont {Koch}\ \emph {et~al.}(2010)\citenamefont {Koch},
  \citenamefont {Sames}, \citenamefont {Kubanek}, \citenamefont {Apel},
  \citenamefont {Balbach}, \citenamefont {Ourjoumtsev}, \citenamefont
  {Pinkse},\ and\ \citenamefont {Rempe}}]{Koch:2010}%
  \BibitemOpen
  \bibfield  {author} {\bibinfo {author} {\bibfnamefont {M.}~\bibnamefont
  {Koch}}, \bibinfo {author} {\bibfnamefont {C.}~\bibnamefont {Sames}},
  \bibinfo {author} {\bibfnamefont {A.}~\bibnamefont {Kubanek}}, \bibinfo
  {author} {\bibfnamefont {M.}~\bibnamefont {Apel}}, \bibinfo {author}
  {\bibfnamefont {M.}~\bibnamefont {Balbach}}, \bibinfo {author} {\bibfnamefont
  {A.}~\bibnamefont {Ourjoumtsev}}, \bibinfo {author} {\bibfnamefont
  {P.}~\bibnamefont {Pinkse}}, \ and\ \bibinfo {author} {\bibfnamefont
  {G.}~\bibnamefont {Rempe}},\ }\href {\doibase 10.1103/PhysRevLett.105.173003}
  {\bibfield  {journal} {\bibinfo  {journal} {Phys. Rev. Lett.}\ }\textbf
  {\bibinfo {volume} {105}},\ \bibinfo {pages} {173003} (\bibinfo {year}
  {2010})}\BibitemShut {NoStop}%
\bibitem [{\citenamefont {Savard}\ \emph {et~al.}(1997)\citenamefont {Savard},
  \citenamefont {O'Hara},\ and\ \citenamefont {Thomas}}]{Savard:1997}%
  \BibitemOpen
  \bibfield  {author} {\bibinfo {author} {\bibfnamefont {T.}~\bibnamefont
  {Savard}}, \bibinfo {author} {\bibfnamefont {K.}~\bibnamefont {O'Hara}}, \
  and\ \bibinfo {author} {\bibfnamefont {J.}~\bibnamefont {Thomas}},\ }\href
  {\doibase 10.1103/PhysRevA.56.R1095} {\bibfield  {journal} {\bibinfo
  {journal} {Phys. Rev. A}\ }\textbf {\bibinfo {volume} {56}},\ \bibinfo
  {pages} {R1095} (\bibinfo {year} {1997})}\BibitemShut {NoStop}%
\bibitem [{\citenamefont {Gehm}\ \emph {et~al.}(1998)\citenamefont {Gehm},
  \citenamefont {O'Hara}, \citenamefont {Savard},\ and\ \citenamefont
  {Thomas}}]{Gehm:1998}%
  \BibitemOpen
  \bibfield  {author} {\bibinfo {author} {\bibfnamefont {M.}~\bibnamefont
  {Gehm}}, \bibinfo {author} {\bibfnamefont {K.}~\bibnamefont {O'Hara}},
  \bibinfo {author} {\bibfnamefont {T.}~\bibnamefont {Savard}}, \ and\ \bibinfo
  {author} {\bibfnamefont {J.}~\bibnamefont {Thomas}},\ }\href {\doibase
  10.1103/PhysRevA.58.3914} {\bibfield  {journal} {\bibinfo  {journal} {Phys.
  Rev. A}\ }\textbf {\bibinfo {volume} {58}},\ \bibinfo {pages} {3914}
  (\bibinfo {year} {1998})}\BibitemShut {NoStop}%
\bibitem [{\citenamefont {Friebel}\ \emph {et~al.}(1998)\citenamefont
  {Friebel}, \citenamefont {D'Andrea}, \citenamefont {Walz}, \citenamefont
  {Weitz},\ and\ \citenamefont {H\"ansch}}]{Friebel:1998}%
  \BibitemOpen
  \bibfield  {author} {\bibinfo {author} {\bibfnamefont {S.}~\bibnamefont
  {Friebel}}, \bibinfo {author} {\bibfnamefont {C.}~\bibnamefont {D'Andrea}},
  \bibinfo {author} {\bibfnamefont {J.}~\bibnamefont {Walz}}, \bibinfo {author}
  {\bibfnamefont {M.}~\bibnamefont {Weitz}}, \ and\ \bibinfo {author}
  {\bibfnamefont {T.~W.}\ \bibnamefont {H\"ansch}},\ }\href {\doibase
  10.1103/PhysRevA.57.R20} {\bibfield  {journal} {\bibinfo  {journal} {Phys.
  Rev. A}\ }\textbf {\bibinfo {volume} {57}},\ \bibinfo {pages} {R20} (\bibinfo
  {year} {1998})}\BibitemShut {NoStop}%
\bibitem [{\citenamefont {Poli}\ \emph {et~al.}(2002)\citenamefont {Poli},
  \citenamefont {Brecha}, \citenamefont {Roati},\ and\ \citenamefont
  {Modugno}}]{Poli:2002}%
  \BibitemOpen
  \bibfield  {author} {\bibinfo {author} {\bibfnamefont {N.}~\bibnamefont
  {Poli}}, \bibinfo {author} {\bibfnamefont {R.}~\bibnamefont {Brecha}},
  \bibinfo {author} {\bibfnamefont {G.}~\bibnamefont {Roati}}, \ and\ \bibinfo
  {author} {\bibfnamefont {G.}~\bibnamefont {Modugno}},\ }\href {\doibase
  10.1103/PhysRevA.65.021401} {\bibfield  {journal} {\bibinfo  {journal} {Phys.
  Rev. A}\ }\textbf {\bibinfo {volume} {65}},\ \bibinfo {pages} {021401}
  (\bibinfo {year} {2002})}\BibitemShut {NoStop}%
\bibitem [{\citenamefont {Turner}\ \emph {et~al.}(1998)\citenamefont {Turner},
  \citenamefont {Miller}, \citenamefont {Hartwell}, \citenamefont {Macdonald},
  \citenamefont {Strogatz},\ and\ \citenamefont {Adams}}]{Turner:1998}%
  \BibitemOpen
  \bibfield  {author} {\bibinfo {author} {\bibfnamefont {K.~L.}\ \bibnamefont
  {Turner}}, \bibinfo {author} {\bibfnamefont {S.~A.}\ \bibnamefont {Miller}},
  \bibinfo {author} {\bibfnamefont {P.~G.}\ \bibnamefont {Hartwell}}, \bibinfo
  {author} {\bibfnamefont {N.~C.}\ \bibnamefont {Macdonald}}, \bibinfo {author}
  {\bibfnamefont {S.~H.}\ \bibnamefont {Strogatz}}, \ and\ \bibinfo {author}
  {\bibfnamefont {S.~G.}\ \bibnamefont {Adams}},\ }\href@noop {} {\bibfield
  {journal} {\bibinfo  {journal} {Nature}\ }\textbf {\bibinfo {volume} {396}},\
  \bibinfo {pages} {149} (\bibinfo {year} {1998})}\BibitemShut {NoStop}%
\bibitem [{\citenamefont {Villanueva}\ \emph {et~al.}(2011)\citenamefont
  {Villanueva}, \citenamefont {Karabalin}, \citenamefont {Matheny},
  \citenamefont {Kenig}, \citenamefont {Cross},\ and\ \citenamefont
  {Roukes}}]{Villanueva:2011}%
  \BibitemOpen
  \bibfield  {author} {\bibinfo {author} {\bibfnamefont {L.~G.}\ \bibnamefont
  {Villanueva}}, \bibinfo {author} {\bibfnamefont {R.~B.}\ \bibnamefont
  {Karabalin}}, \bibinfo {author} {\bibfnamefont {M.~H.}\ \bibnamefont
  {Matheny}}, \bibinfo {author} {\bibfnamefont {E.}~\bibnamefont {Kenig}},
  \bibinfo {author} {\bibfnamefont {M.~C.}\ \bibnamefont {Cross}}, \ and\
  \bibinfo {author} {\bibfnamefont {M.~L.}\ \bibnamefont {Roukes}},\ }\href
  {\doibase 10.1021/nl2031162} {\bibfield  {journal} {\bibinfo  {journal} {Nano
  Lett.}\ }\textbf {\bibinfo {volume} {11}},\ \bibinfo {pages} {5054} (\bibinfo
  {year} {2011})}\BibitemShut {NoStop}%
\bibitem [{\citenamefont {Vinante}\ and\ \citenamefont
  {Falferi}(2013)}]{Vinante:2013}%
  \BibitemOpen
  \bibfield  {author} {\bibinfo {author} {\bibfnamefont {A.}~\bibnamefont
  {Vinante}}\ and\ \bibinfo {author} {\bibfnamefont {P.}~\bibnamefont
  {Falferi}},\ }\href {\doibase 10.1103/PhysRevLett.111.207203} {\bibfield
  {journal} {\bibinfo  {journal} {Phys. Rev. Lett.}\ }\textbf {\bibinfo
  {volume} {111}},\ \bibinfo {pages} {207203} (\bibinfo {year}
  {2013})}\BibitemShut {NoStop}%
\bibitem [{\citenamefont {Gieseler}\ \emph {et~al.}(2012)\citenamefont
  {Gieseler}, \citenamefont {Deutsch}, \citenamefont {Quidant},\ and\
  \citenamefont {Novotny}}]{Gieseler:2012}%
  \BibitemOpen
  \bibfield  {author} {\bibinfo {author} {\bibfnamefont {J.}~\bibnamefont
  {Gieseler}}, \bibinfo {author} {\bibfnamefont {B.}~\bibnamefont {Deutsch}},
  \bibinfo {author} {\bibfnamefont {R.}~\bibnamefont {Quidant}}, \ and\
  \bibinfo {author} {\bibfnamefont {L.}~\bibnamefont {Novotny}},\ }\href
  {\doibase 10.1103/PhysRevLett.109.103603} {\bibfield  {journal} {\bibinfo
  {journal} {Phys. Rev. Lett.}\ }\textbf {\bibinfo {volume} {109}},\ \bibinfo
  {pages} {103603} (\bibinfo {year} {2012})}\BibitemShut {NoStop}%
\bibitem [{\citenamefont {Jain}\ \emph {et~al.}(2016)\citenamefont {Jain},
  \citenamefont {Gieseler}, \citenamefont {Moritz}, \citenamefont {Dellago},
  \citenamefont {Quidant},\ and\ \citenamefont {Novotny}}]{Jain:2016}%
  \BibitemOpen
  \bibfield  {author} {\bibinfo {author} {\bibfnamefont {V.}~\bibnamefont
  {Jain}}, \bibinfo {author} {\bibfnamefont {J.}~\bibnamefont {Gieseler}},
  \bibinfo {author} {\bibfnamefont {C.}~\bibnamefont {Moritz}}, \bibinfo
  {author} {\bibfnamefont {C.}~\bibnamefont {Dellago}}, \bibinfo {author}
  {\bibfnamefont {R.}~\bibnamefont {Quidant}}, \ and\ \bibinfo {author}
  {\bibfnamefont {L.}~\bibnamefont {Novotny}},\ }\href {\doibase
  10.1103/PhysRevLett.116.243601} {\bibfield  {journal} {\bibinfo  {journal}
  {Phys. Rev. Lett.}\ }\textbf {\bibinfo {volume} {116}},\ \bibinfo {pages}
  {243601} (\bibinfo {year} {2016})}\BibitemShut {NoStop}%
\bibitem [{\citenamefont {Bushev}\ \emph {et~al.}(2006)\citenamefont {Bushev},
  \citenamefont {Rotter}, \citenamefont {Wilson}, \citenamefont {Dubin},
  \citenamefont {Becher}, \citenamefont {Eschner}, \citenamefont {Blatt},
  \citenamefont {Steixner}, \citenamefont {Rabl},\ and\ \citenamefont
  {Zoller}}]{Bushev:2006}%
  \BibitemOpen
  \bibfield  {author} {\bibinfo {author} {\bibfnamefont {P.}~\bibnamefont
  {Bushev}}, \bibinfo {author} {\bibfnamefont {D.}~\bibnamefont {Rotter}},
  \bibinfo {author} {\bibfnamefont {A.}~\bibnamefont {Wilson}}, \bibinfo
  {author} {\bibfnamefont {F.}~\bibnamefont {Dubin}}, \bibinfo {author}
  {\bibfnamefont {C.}~\bibnamefont {Becher}}, \bibinfo {author} {\bibfnamefont
  {J.}~\bibnamefont {Eschner}}, \bibinfo {author} {\bibfnamefont
  {R.}~\bibnamefont {Blatt}}, \bibinfo {author} {\bibfnamefont
  {V.}~\bibnamefont {Steixner}}, \bibinfo {author} {\bibfnamefont
  {P.}~\bibnamefont {Rabl}}, \ and\ \bibinfo {author} {\bibfnamefont
  {P.}~\bibnamefont {Zoller}},\ }\href {\doibase 10.1103/PhysRevLett.96.043003}
  {\bibfield  {journal} {\bibinfo  {journal} {Phys. Rev. Lett.}\ }\textbf
  {\bibinfo {volume} {96}},\ \bibinfo {pages} {043003} (\bibinfo {year}
  {2006})}\BibitemShut {NoStop}%
\bibitem [{\citenamefont {Koch}\ \emph {et~al.}(2011)\citenamefont {Koch},
  \citenamefont {Sames}, \citenamefont {Balbach}, \citenamefont {Chibani},
  \citenamefont {Kubanek}, \citenamefont {Murr}, \citenamefont {Wilk},\ and\
  \citenamefont {Rempe}}]{Koch:2011}%
  \BibitemOpen
  \bibfield  {author} {\bibinfo {author} {\bibfnamefont {M.}~\bibnamefont
  {Koch}}, \bibinfo {author} {\bibfnamefont {C.}~\bibnamefont {Sames}},
  \bibinfo {author} {\bibfnamefont {M.}~\bibnamefont {Balbach}}, \bibinfo
  {author} {\bibfnamefont {H.}~\bibnamefont {Chibani}}, \bibinfo {author}
  {\bibfnamefont {A.}~\bibnamefont {Kubanek}}, \bibinfo {author} {\bibfnamefont
  {K.}~\bibnamefont {Murr}}, \bibinfo {author} {\bibfnamefont {T.}~\bibnamefont
  {Wilk}}, \ and\ \bibinfo {author} {\bibfnamefont {G.}~\bibnamefont {Rempe}},\
  }\href {\doibase 10.1103/PhysRevLett.107.023601} {\bibfield  {journal}
  {\bibinfo  {journal} {Phys. Rev. Lett.}\ }\textbf {\bibinfo {volume} {107}},\
  \bibinfo {pages} {2} (\bibinfo {year} {2011})}\BibitemShut {NoStop}%
\bibitem [{\citenamefont {Boca}\ \emph {et~al.}(2004)\citenamefont {Boca},
  \citenamefont {Miller}, \citenamefont {Birnbaum}, \citenamefont {Boozer},
  \citenamefont {McKeever},\ and\ \citenamefont {Kimble}}]{Boca:2004}%
  \BibitemOpen
  \bibfield  {author} {\bibinfo {author} {\bibfnamefont {A.}~\bibnamefont
  {Boca}}, \bibinfo {author} {\bibfnamefont {R.}~\bibnamefont {Miller}},
  \bibinfo {author} {\bibfnamefont {K.~M.}\ \bibnamefont {Birnbaum}}, \bibinfo
  {author} {\bibfnamefont {A.~D.}\ \bibnamefont {Boozer}}, \bibinfo {author}
  {\bibfnamefont {J.}~\bibnamefont {McKeever}}, \ and\ \bibinfo {author}
  {\bibfnamefont {H.~J.}\ \bibnamefont {Kimble}},\ }\href {\doibase
  10.1103/PhysRevLett.93.233603} {\bibfield  {journal} {\bibinfo  {journal}
  {Phys. Rev. Lett.}\ }\textbf {\bibinfo {volume} {93}},\ \bibinfo {pages}
  {233603} (\bibinfo {year} {2004})}\BibitemShut {NoStop}%
\bibitem [{\citenamefont {Maunz}\ \emph {et~al.}(2005)\citenamefont {Maunz},
  \citenamefont {Puppe}, \citenamefont {Schuster}, \citenamefont {Syassen},
  \citenamefont {Pinkse},\ and\ \citenamefont {Rempe}}]{Maunz:2005}%
  \BibitemOpen
  \bibfield  {author} {\bibinfo {author} {\bibfnamefont {P.}~\bibnamefont
  {Maunz}}, \bibinfo {author} {\bibfnamefont {T.}~\bibnamefont {Puppe}},
  \bibinfo {author} {\bibfnamefont {I.}~\bibnamefont {Schuster}}, \bibinfo
  {author} {\bibfnamefont {N.}~\bibnamefont {Syassen}}, \bibinfo {author}
  {\bibfnamefont {P.~W.~H.}\ \bibnamefont {Pinkse}}, \ and\ \bibinfo {author}
  {\bibfnamefont {G.}~\bibnamefont {Rempe}},\ }\href {\doibase
  10.1103/PhysRevLett.94.033002} {\bibfield  {journal} {\bibinfo  {journal}
  {Phys. Rev. Lett.}\ }\textbf {\bibinfo {volume} {94}},\ \bibinfo {pages}
  {033002} (\bibinfo {year} {2005})}\BibitemShut {NoStop}%
\bibitem [{\citenamefont {Kubanek}\ \emph {et~al.}(2011)\citenamefont
  {Kubanek}, \citenamefont {Koch}, \citenamefont {Sames}, \citenamefont
  {Ourjoumtsev}, \citenamefont {Wilk}, \citenamefont {Pinkse},\ and\
  \citenamefont {Rempe}}]{Kubanek:2011}%
  \BibitemOpen
  \bibfield  {author} {\bibinfo {author} {\bibfnamefont {A.}~\bibnamefont
  {Kubanek}}, \bibinfo {author} {\bibfnamefont {M.}~\bibnamefont {Koch}},
  \bibinfo {author} {\bibfnamefont {C.}~\bibnamefont {Sames}}, \bibinfo
  {author} {\bibfnamefont {A.}~\bibnamefont {Ourjoumtsev}}, \bibinfo {author}
  {\bibfnamefont {T.}~\bibnamefont {Wilk}}, \bibinfo {author} {\bibfnamefont
  {P.~W.~H.}\ \bibnamefont {Pinkse}}, \ and\ \bibinfo {author} {\bibfnamefont
  {G.}~\bibnamefont {Rempe}},\ }\href {\doibase 10.1007/s00340-011-4410-x}
  {\bibfield  {journal} {\bibinfo  {journal} {Applied Physics B}\ }\textbf
  {\bibinfo {volume} {102}},\ \bibinfo {pages} {433} (\bibinfo {year}
  {2011})}\BibitemShut {NoStop}%
\bibitem [{\citenamefont {Hamsen}\ \emph {et~al.}(2017)\citenamefont {Hamsen},
  \citenamefont {Tolazzi}, \citenamefont {Wilk},\ and\ \citenamefont
  {Rempe}}]{Hamsen:2017}%
  \BibitemOpen
  \bibfield  {author} {\bibinfo {author} {\bibfnamefont {C.}~\bibnamefont
  {Hamsen}}, \bibinfo {author} {\bibfnamefont {K.~N.}\ \bibnamefont {Tolazzi}},
  \bibinfo {author} {\bibfnamefont {T.}~\bibnamefont {Wilk}}, \ and\ \bibinfo
  {author} {\bibfnamefont {G.}~\bibnamefont {Rempe}},\ }\href {\doibase
  10.1103/PhysRevLett.118.133604} {\bibfield  {journal} {\bibinfo  {journal}
  {Phys. Rev. Lett.}\ }\textbf {\bibinfo {volume} {118}},\ \bibinfo {pages}
  {133604} (\bibinfo {year} {2017})}\BibitemShut {NoStop}%
\bibitem [{\citenamefont {Turchette}\ \emph {et~al.}(1995)\citenamefont
  {Turchette}, \citenamefont {Hood}, \citenamefont {Lange}, \citenamefont
  {Mabuchi},\ and\ \citenamefont {Kimble}}]{Turchette:1995}%
  \BibitemOpen
  \bibfield  {author} {\bibinfo {author} {\bibfnamefont {Q.~A.}\ \bibnamefont
  {Turchette}}, \bibinfo {author} {\bibfnamefont {C.~J.}\ \bibnamefont {Hood}},
  \bibinfo {author} {\bibfnamefont {W.}~\bibnamefont {Lange}}, \bibinfo
  {author} {\bibfnamefont {H.}~\bibnamefont {Mabuchi}}, \ and\ \bibinfo
  {author} {\bibfnamefont {H.~J.}\ \bibnamefont {Kimble}},\ }\href {\doibase
  10.1103/PhysRevLett.75.4710} {\bibfield  {journal} {\bibinfo  {journal}
  {Phys. Rev. Lett.}\ }\textbf {\bibinfo {volume} {75}},\ \bibinfo {pages}
  {4710} (\bibinfo {year} {1995})}\BibitemShut {NoStop}%
\bibitem [{\citenamefont {Beck}\ \emph {et~al.}(2016)\citenamefont {Beck},
  \citenamefont {Hosseini}, \citenamefont {Duan},\ and\ \citenamefont
  {Vuleti{\'c}}}]{Beck:2016}%
  \BibitemOpen
  \bibfield  {author} {\bibinfo {author} {\bibfnamefont {K.~M.}\ \bibnamefont
  {Beck}}, \bibinfo {author} {\bibfnamefont {M.}~\bibnamefont {Hosseini}},
  \bibinfo {author} {\bibfnamefont {Y.}~\bibnamefont {Duan}}, \ and\ \bibinfo
  {author} {\bibfnamefont {V.}~\bibnamefont {Vuleti{\'c}}},\ }\href {\doibase
  10.1073/pnas.1524117113} {\bibfield  {journal} {\bibinfo  {journal}
  {Proceedings of the National Academy of Sciences}\ }\textbf {\bibinfo
  {volume} {113}},\ \bibinfo {pages} {9740} (\bibinfo {year}
  {2016})}\BibitemShut {NoStop}%
\bibitem [{\citenamefont {Northup}\ and\ \citenamefont
  {Blatt}(2014)}]{Northup:2014}%
  \BibitemOpen
  \bibfield  {author} {\bibinfo {author} {\bibfnamefont {T.~E.}\ \bibnamefont
  {Northup}}\ and\ \bibinfo {author} {\bibfnamefont {R.}~\bibnamefont
  {Blatt}},\ }\href {http://dx.doi.org/10.1038/nphoton.2014.53} {\bibfield
  {journal} {\bibinfo  {journal} {Nature Photonics}\ }\textbf {\bibinfo
  {volume} {8}},\ \bibinfo {pages} {356 EP } (\bibinfo {year}
  {2014})}\BibitemShut {NoStop}%
\bibitem [{\citenamefont {Reiserer}\ and\ \citenamefont
  {Rempe}(2015)}]{Reiserer:2015}%
  \BibitemOpen
  \bibfield  {author} {\bibinfo {author} {\bibfnamefont {A.}~\bibnamefont
  {Reiserer}}\ and\ \bibinfo {author} {\bibfnamefont {G.}~\bibnamefont
  {Rempe}},\ }\href {\doibase 10.1103/RevModPhys.87.1379} {\bibfield  {journal}
  {\bibinfo  {journal} {Rev. Mod. Phys.}\ }\textbf {\bibinfo {volume} {87}},\
  \bibinfo {pages} {1379} (\bibinfo {year} {2015})}\BibitemShut {NoStop}%
\bibitem [{\citenamefont {Wilk}\ \emph {et~al.}(2007)\citenamefont {Wilk},
  \citenamefont {Webster}, \citenamefont {Kuhn},\ and\ \citenamefont
  {Rempe}}]{Wilk:2007}%
  \BibitemOpen
  \bibfield  {author} {\bibinfo {author} {\bibfnamefont {T.}~\bibnamefont
  {Wilk}}, \bibinfo {author} {\bibfnamefont {S.~C.}\ \bibnamefont {Webster}},
  \bibinfo {author} {\bibfnamefont {A.}~\bibnamefont {Kuhn}}, \ and\ \bibinfo
  {author} {\bibfnamefont {G.}~\bibnamefont {Rempe}},\ }\href {\doibase
  10.1126/science.1143835} {\bibfield  {journal} {\bibinfo  {journal}
  {Science}\ }\textbf {\bibinfo {volume} {317}},\ \bibinfo {pages} {488}
  (\bibinfo {year} {2007})}\BibitemShut {NoStop}%
\bibitem [{\citenamefont {Weber}\ \emph {et~al.}(2009)\citenamefont {Weber},
  \citenamefont {Specht}, \citenamefont {M\"uller}, \citenamefont {Bochmann},
  \citenamefont {M\"ucke}, \citenamefont {Moehring},\ and\ \citenamefont
  {Rempe}}]{Weber:2009}%
  \BibitemOpen
  \bibfield  {author} {\bibinfo {author} {\bibfnamefont {B.}~\bibnamefont
  {Weber}}, \bibinfo {author} {\bibfnamefont {H.~P.}\ \bibnamefont {Specht}},
  \bibinfo {author} {\bibfnamefont {T.}~\bibnamefont {M\"uller}}, \bibinfo
  {author} {\bibfnamefont {J.}~\bibnamefont {Bochmann}}, \bibinfo {author}
  {\bibfnamefont {M.}~\bibnamefont {M\"ucke}}, \bibinfo {author} {\bibfnamefont
  {D.~L.}\ \bibnamefont {Moehring}}, \ and\ \bibinfo {author} {\bibfnamefont
  {G.}~\bibnamefont {Rempe}},\ }\href {\doibase 10.1103/PhysRevLett.102.030501}
  {\bibfield  {journal} {\bibinfo  {journal} {Phys. Rev. Lett.}\ }\textbf
  {\bibinfo {volume} {102}},\ \bibinfo {pages} {030501} (\bibinfo {year}
  {2009})}\BibitemShut {NoStop}%
\bibitem [{\citenamefont {Stute}\ \emph {et~al.}(2012)\citenamefont {Stute},
  \citenamefont {Casabone}, \citenamefont {Schindler}, \citenamefont {Monz},
  \citenamefont {Schmidt}, \citenamefont {Brandst{\"a}tter}, \citenamefont
  {Northup},\ and\ \citenamefont {Blatt}}]{Stute:2012}%
  \BibitemOpen
  \bibfield  {author} {\bibinfo {author} {\bibfnamefont {A.}~\bibnamefont
  {Stute}}, \bibinfo {author} {\bibfnamefont {B.}~\bibnamefont {Casabone}},
  \bibinfo {author} {\bibfnamefont {P.}~\bibnamefont {Schindler}}, \bibinfo
  {author} {\bibfnamefont {T.}~\bibnamefont {Monz}}, \bibinfo {author}
  {\bibfnamefont {P.~O.}\ \bibnamefont {Schmidt}}, \bibinfo {author}
  {\bibfnamefont {B.}~\bibnamefont {Brandst{\"a}tter}}, \bibinfo {author}
  {\bibfnamefont {T.~E.}\ \bibnamefont {Northup}}, \ and\ \bibinfo {author}
  {\bibfnamefont {R.}~\bibnamefont {Blatt}},\ }\href
  {http://dx.doi.org/10.1038/nature11120} {\bibfield  {journal} {\bibinfo
  {journal} {Nature}\ }\textbf {\bibinfo {volume} {485}},\ \bibinfo {pages}
  {482 EP } (\bibinfo {year} {2012})}\BibitemShut {NoStop}%
\bibitem [{\citenamefont {Specht}\ \emph {et~al.}(2011)\citenamefont {Specht},
  \citenamefont {N{\"o}lleke}, \citenamefont {Reiserer}, \citenamefont
  {Uphoff}, \citenamefont {Figueroa}, \citenamefont {Ritter},\ and\
  \citenamefont {Rempe}}]{Specht:2011}%
  \BibitemOpen
  \bibfield  {author} {\bibinfo {author} {\bibfnamefont {H.~P.}\ \bibnamefont
  {Specht}}, \bibinfo {author} {\bibfnamefont {C.}~\bibnamefont {N{\"o}lleke}},
  \bibinfo {author} {\bibfnamefont {A.}~\bibnamefont {Reiserer}}, \bibinfo
  {author} {\bibfnamefont {M.}~\bibnamefont {Uphoff}}, \bibinfo {author}
  {\bibfnamefont {E.}~\bibnamefont {Figueroa}}, \bibinfo {author}
  {\bibfnamefont {S.}~\bibnamefont {Ritter}}, \ and\ \bibinfo {author}
  {\bibfnamefont {G.}~\bibnamefont {Rempe}},\ }\href
  {http://dx.doi.org/10.1038/nature09997} {\bibfield  {journal} {\bibinfo
  {journal} {Nature}\ }\textbf {\bibinfo {volume} {473}},\ \bibinfo {pages}
  {190 EP } (\bibinfo {year} {2011})}\BibitemShut {NoStop}%
\bibitem [{\citenamefont {K{\"o}rber}\ \emph {et~al.}(2018)\citenamefont
  {K{\"o}rber}, \citenamefont {Morin}, \citenamefont {Langenfeld},
  \citenamefont {Neuzner}, \citenamefont {Ritter},\ and\ \citenamefont
  {Rempe}}]{Korber:2018}%
  \BibitemOpen
  \bibfield  {author} {\bibinfo {author} {\bibfnamefont {M.}~\bibnamefont
  {K{\"o}rber}}, \bibinfo {author} {\bibfnamefont {O.}~\bibnamefont {Morin}},
  \bibinfo {author} {\bibfnamefont {S.}~\bibnamefont {Langenfeld}}, \bibinfo
  {author} {\bibfnamefont {A.}~\bibnamefont {Neuzner}}, \bibinfo {author}
  {\bibfnamefont {S.}~\bibnamefont {Ritter}}, \ and\ \bibinfo {author}
  {\bibfnamefont {G.}~\bibnamefont {Rempe}},\ }\href {\doibase
  10.1038/s41566-017-0050-y} {\bibfield  {journal} {\bibinfo  {journal} {Nature
  Photonics}\ }\textbf {\bibinfo {volume} {12}},\ \bibinfo {pages} {18}
  (\bibinfo {year} {2018})}\BibitemShut {NoStop}%
\bibitem [{\citenamefont {Dayan}\ \emph {et~al.}(2008)\citenamefont {Dayan},
  \citenamefont {Parkins}, \citenamefont {Aoki}, \citenamefont {Ostby},
  \citenamefont {Vahala},\ and\ \citenamefont {Kimble}}]{Dayan:2008}%
  \BibitemOpen
  \bibfield  {author} {\bibinfo {author} {\bibfnamefont {B.}~\bibnamefont
  {Dayan}}, \bibinfo {author} {\bibfnamefont {A.~S.}\ \bibnamefont {Parkins}},
  \bibinfo {author} {\bibfnamefont {T.}~\bibnamefont {Aoki}}, \bibinfo {author}
  {\bibfnamefont {E.~P.}\ \bibnamefont {Ostby}}, \bibinfo {author}
  {\bibfnamefont {K.~J.}\ \bibnamefont {Vahala}}, \ and\ \bibinfo {author}
  {\bibfnamefont {H.~J.}\ \bibnamefont {Kimble}},\ }\href {\doibase
  10.1126/science.1152261} {\bibfield  {journal} {\bibinfo  {journal}
  {Science}\ }\textbf {\bibinfo {volume} {319}},\ \bibinfo {pages} {1062}
  (\bibinfo {year} {2008})}\BibitemShut {NoStop}%
\bibitem [{\citenamefont {Gehr}\ \emph {et~al.}(2010)\citenamefont {Gehr},
  \citenamefont {Volz}, \citenamefont {Dubois}, \citenamefont {Steinmetz},
  \citenamefont {Colombe}, \citenamefont {Lev}, \citenamefont {Long},
  \citenamefont {Est\`eve},\ and\ \citenamefont {Reichel}}]{Gehr:2010}%
  \BibitemOpen
  \bibfield  {author} {\bibinfo {author} {\bibfnamefont {R.}~\bibnamefont
  {Gehr}}, \bibinfo {author} {\bibfnamefont {J.}~\bibnamefont {Volz}}, \bibinfo
  {author} {\bibfnamefont {G.}~\bibnamefont {Dubois}}, \bibinfo {author}
  {\bibfnamefont {T.}~\bibnamefont {Steinmetz}}, \bibinfo {author}
  {\bibfnamefont {Y.}~\bibnamefont {Colombe}}, \bibinfo {author} {\bibfnamefont
  {B.~L.}\ \bibnamefont {Lev}}, \bibinfo {author} {\bibfnamefont
  {R.}~\bibnamefont {Long}}, \bibinfo {author} {\bibfnamefont {J.}~\bibnamefont
  {Est\`eve}}, \ and\ \bibinfo {author} {\bibfnamefont {J.}~\bibnamefont
  {Reichel}},\ }\href {\doibase 10.1103/PhysRevLett.104.203602} {\bibfield
  {journal} {\bibinfo  {journal} {Phys. Rev. Lett.}\ }\textbf {\bibinfo
  {volume} {104}},\ \bibinfo {pages} {203602} (\bibinfo {year}
  {2010})}\BibitemShut {NoStop}%
\bibitem [{\citenamefont {Rosenblum}\ \emph {et~al.}(2015)\citenamefont
  {Rosenblum}, \citenamefont {Bechler}, \citenamefont {Shomroni}, \citenamefont
  {Lovsky}, \citenamefont {Guendelman},\ and\ \citenamefont
  {Dayan}}]{Rosenblum:2015}%
  \BibitemOpen
  \bibfield  {author} {\bibinfo {author} {\bibfnamefont {S.}~\bibnamefont
  {Rosenblum}}, \bibinfo {author} {\bibfnamefont {O.}~\bibnamefont {Bechler}},
  \bibinfo {author} {\bibfnamefont {I.}~\bibnamefont {Shomroni}}, \bibinfo
  {author} {\bibfnamefont {Y.}~\bibnamefont {Lovsky}}, \bibinfo {author}
  {\bibfnamefont {G.}~\bibnamefont {Guendelman}}, \ and\ \bibinfo {author}
  {\bibfnamefont {B.}~\bibnamefont {Dayan}},\ }\href
  {http://dx.doi.org/10.1038/nphoton.2015.227} {\bibfield  {journal} {\bibinfo
  {journal} {Nature Photonics}\ }\textbf {\bibinfo {volume} {10}},\ \bibinfo
  {pages} {19} (\bibinfo {year} {2015})}\BibitemShut {NoStop}%
\bibitem [{\citenamefont {Scheucher}\ \emph {et~al.}(2016)\citenamefont
  {Scheucher}, \citenamefont {Hilico}, \citenamefont {Will}, \citenamefont
  {Volz},\ and\ \citenamefont {Rauschenbeutel}}]{Scheucher:2016}%
  \BibitemOpen
  \bibfield  {author} {\bibinfo {author} {\bibfnamefont {M.}~\bibnamefont
  {Scheucher}}, \bibinfo {author} {\bibfnamefont {A.}~\bibnamefont {Hilico}},
  \bibinfo {author} {\bibfnamefont {E.}~\bibnamefont {Will}}, \bibinfo {author}
  {\bibfnamefont {J.}~\bibnamefont {Volz}}, \ and\ \bibinfo {author}
  {\bibfnamefont {A.}~\bibnamefont {Rauschenbeutel}},\ }\href {\doibase
  10.1126/science.aaj2118} {\bibfield  {journal} {\bibinfo  {journal}
  {Science}\ } (\bibinfo {year} {2016}),\ 10.1126/science.aaj2118}\BibitemShut
  {NoStop}%
\end{thebibliography}%

\end{document}